# Dimensionless solutions and general characteristics of bioheat transfer during thermal therapy

Junnosuke Okajima[1], Shigenao Maruyama[2], Hiroki Takeda[1] and Atsuki Komiya[2]

[1]Graduate School of Engineering, Tohoku University, 6-6-04, Aramaki Aza Aoba Aoba-ku, Sendai, Miyagi 980-8579, Japan.

[2]Institute of Fluid Science, Tohoku University, 2-1-1, Katahira, Aoba-ku, Sendai, Miyagi 980-8577, Japan.

The derivation and application of the general characteristics of bioheat transfer for medical applications are shown in this paper. Two general bioheat transfer characteristics are derived from solutions of one-dimensional Pennes' bioheat transfer equation: steady-state thermal penetration depth, which is the deepest depth where the heat effect reaches; and time to reach steady state, which represents the amount of time necessary for temperature distribution to converge to a steady-state. All results are described by dimensionless form; therefore these results provide information on temperature distribution in biological tissue for various thermal therapies by transforming to dimension form.

## 1. Introduction

Clinicians make use of a number of treatments that utilize heat transfer phenomena (Diller and Ryan, 1998). These thermal therapies affect biological tissues by changing their temperature. Such methods include moxibustion, which is local heating method using combustion of moxa (herb) and developed in oriental medicine (Shen et al., 2006); hypothermia, which inhibits the metabolism by cooling (Ji and Liu, 2002; Diao et al., 2003); and hyperthermia, which heats the tissue locally by electromagnetic waves (Brix et al., 2002; Thiebaut and Lemonnier, 2002). The thermal responses of the living body under these treatments have not yet been fully evaluated quantitatively in the clinical field because these treatments have depended on the doctor's experience.

Meanwhile it is difficult to analyze the thermal response in biological tissues precisely because of the mechanisms that maintain body temperature, such as blood flow and metabolic heat generation. Therefore, many different bioheat transfer models have been proposed (Khaleda and Vafai, 2003). Almost all thermal analyses of thermal therapy consist of a numerical simulation for a specific type of therapy, such as hyperthermia, hypothermia or cryosurgery (Ji and Liu, 2002; Thiebaut and Lemonnier, 2002; Chua et al., 2007). The results thus apply only to the specific treatment and the general characteristics of bioheat transfer

Corresponding author: Junnosuke Okajima

Institute of Fluid Science, Tohoku University, 2-1-1, Katahira, Aoba-ku, Sendai, Miyagi 980-8577, Japan, TEL & FAX: +81-22-217-5879, E-mail: j.okajima@tohoku.ac.jp

during thermal therapy remain to be elucidated.

It is essential to share information on the general characteristics of bioheat transfer with medical doctors in order to investigate the relationship between the effects of treatment and the thermal response of the human body during thermal therapy. However, existing data on thermal analysis are limited to specific treatments and are not extensively available in the clinical field. Furthermore, it is difficult for medical doctors in clinical field to conduct thermal analysis, and it is therefore important to provide doctors with useful data concerning the thermal analysis of biological tissue. The purpose of the present study was thus to examine the analytical solutions of a bioheat transfer equation to derive the general characteristics of bioheat transfer. Some analytical solutions have been derived in previous studies, however they were limited to specific treatments (Brix et al., 2002) or utilized infinite series or special functions complexly (Durkee and Antich, 1991). Brix et al. derived analytical solutions by the Green's function method estimating the temperature rise during MR procedures. However, the discussion was limited to the MR procedure. Furthermore, Durkee and Antich derived transient analytical solutions in a multi-layer structure. The derived solutions were used for the calculation of the temperature distribution in the skin, muscle and bone. Therefore the basic and general bioheat transfer characteristics have not been discussed so far. To obtain the general characteristics of bioheat transfer, we used dimensionless analysis to derive analytical solutions independent of the specific kind of biological tissue or organ. We conducted thermal analyses using dimensionless solutions of the bioheat transfer equation under one-dimensional (1-D) Cartesian coordinates and 1st kind boundary condition (BC) (Maruyama et al., 2008). The solutions correspond to treatments that involve heating/cooling on the surface of biological tissue. Furthermore, we suggest a simple method of estimating temperature distribution in biological tissue by using the bioheat transfer characteristics derived from the solutions.

There are many heating/cooling methods and treatment devices other than heating/cooling by constant temperature on the body surface. For example, we have developed two kinds of heating devices for thermal therapy, one of which utilizes a heated metal disc with a temperature control (Takashima et al., 2008) and can be approximated by 1-D Cartesian coordinates and constant temperature BC, and the other of which utilizes infrared radiation to heat the body surface (Ogasawara et al., 2008) and can be approximated by 1-D Cartesian coordinates and heat flux BC. In addition, body surface cooling with a cooling pad is used in hypothermia (Haugk et al., 2007), and can be approximated by 1-D Cartesian coordinates and convective heat transfer BC, and radiofrequency ablation uses a heating probe and can be approximated by 1-D axisymmetric or spherical symmetric coordinates and heat flux BC. As mentioned above, the coordinates used for thermal analysis change according to the treatment method, as do the BCs. It is important to derive the general characteristics of bioheat transfer that are common to various treatment methods. The objectives of the present study are to derive the dimensionless steady-state solutions of the bioheat transfer equation under various coordinates and BCs and to discuss the bioheat transfer characteristics common to all organs or tissues by using these solutions and several thermophysical properties. A method of estimating temperature distribution in biological tissue is also proposed.

## 2. Models

Figure 1 shows the three kinds of coordinates considered in this study. In Fig. 1, $x_p$ denotes the radius of the treatment probe. The 1-D Cartesian coordinate corresponds to surface heating/cooling. The 1-D axisymmetric coordinate is suitable for treatment using a heating/cooling probe. When the heating/cooling section is small, the 1-D spherical symmetric coordinate is better approximated.

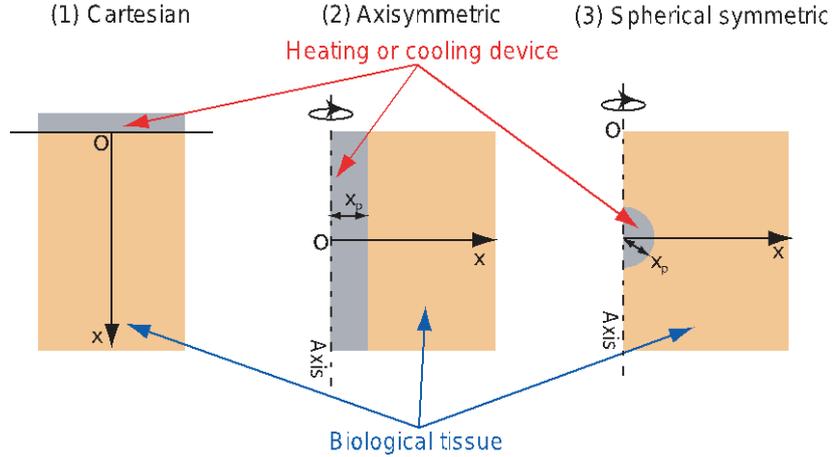

Fig. 1. The models for bioheat transfer analyses at the each coordinate

In the present study, we used Pennes' bioheat transfer equation (Pennes, 1948), which is the simplest and oldest bioheat transfer equation (Charny, 1992). The advantages of this equation are the minimal parameters it uses to describe heat transfer phenomena in biological tissue and the existence of many experimental data about these parameters (Brix et al., 2002; Fiala et al., 1999). The 1-D Pennes' bioheat transfer equation is expressed as

$$\rho c \frac{\partial T}{\partial t} = \frac{k}{x^n} \frac{\partial}{\partial x}\left( x^n \frac{\partial T}{\partial x} \right) + \rho_b c_b \omega_b (T_B - T), \tag{1}$$

where $n$ denotes the classification of the coordinates: $n = 0, 1, 2$ to indicate Cartesian, axisymmteric and spherical symmetric coordinates, respectively. The second term on the right side is the blood perfusion term. $T_B$ is the steady-state temperature of the biological tissue, which is defined as

$$T_B = T_a + \frac{q_{met}}{\rho_b c_b \omega_b}. \tag{2}$$

This temperature is defined by the arterial blood temperature $T_a$ and metabolic heat generation rate $q_{met}$ (Maruyama et al., 2008), and represents the equilibrium temperature of each tissue or organ described by Pennes' bioheat transfer model. For simplicity, the following assumptions are made:

    a. The biological tissue is isotropic and homogeneous.
    b. The tissue properties are independent of the tissue temperature.
    c. The metabolic heat generation rate is constant per unit volume and unit time (Chua et al., 2007) and 4200 W/m$^3$ (Deng and Liu, 2005).
    d. The blood perfusion rate is uniform spatially and temporally and independent of

tissue temperature (Zhang et al., 2005).

e. Arterial blood temperature is constant at 37°C (Zhang et al., 2005).

Furthermore, the steady-state temperature of biological tissue can be used as an initial condition and infinite BC.

$$t = 0 \; ; \; T = T_B, \tag{3}$$

$$x \to \infty \; ; \; T \to T_B. \tag{4}$$

There are three kinds of BCs on the surface of biological tissue (or the boundary between biological tissue and treatment device). Almost all heating/cooling methods include these three kinds of BC, which are expressed as

$$1\text{st kind:} \; T|_{Surface} = T_w, \tag{5}$$

$$2\text{nd kind:} \; -k \frac{\partial T}{\partial x}\bigg|_{Surface} = q_w, \tag{6}$$

$$3\text{rd kind:} \; -k \frac{\partial T}{\partial x}\bigg|_{Surface} = h\left[T_f - T|_{Surface}\right]. \tag{7}$$

First kind BC represents heating/cooling at a constant temperature Tw, second kind BC represents heating/cooling by constant heat flux $q_w$, and third kind BC represents heating/cooling by convective heat transfer, which means heat exchange between the tissue surface and fluid at a constant temperature $T_f$. Here h in Eq. (7) denotes the heat transfer coefficient that represents the ratio of heat exchange.

We validated this model in the case of the 1-D Cartesian coordinate and the first kind BC (Maruyama et al., 2008) using the experimental result of the goat's brain cooling conducted by Maruyama et al. (2006). Figure 2 shows the validation result, which is a comparison between experimental results, numerical simulation results and the analytical solution. This result shows that the analytical solutions of the bioheat transfer equation can be applied in the estimation of the general characteristics of bioheat transfer.

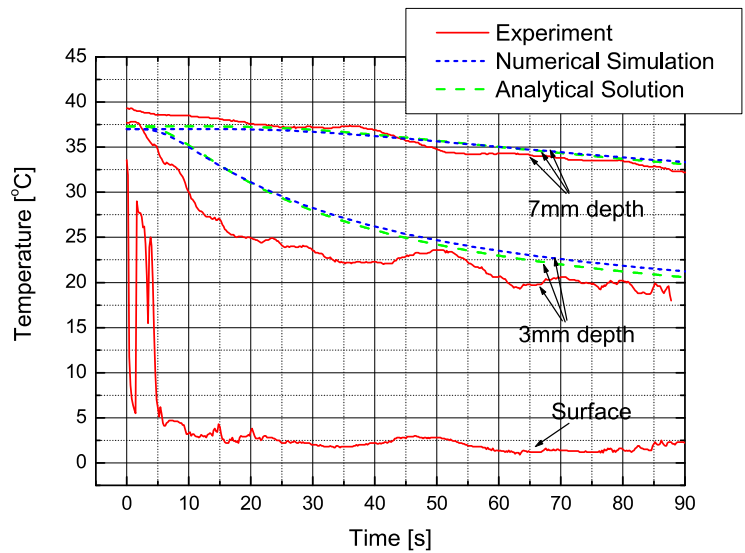

Fig. 2. The comparison among the experimental result of brain cooling, result of numerical simulation and analytical solution.

### 3. Steady-State Solutions

*3.1 Dimensional steady-state solutions*

Pennes' bioheat transfer equation (Eq. (1)) includes the blood perfusion term whose function is to maintain the temperature of biological tissue. Therefore, Eq. (1) has steady-state solutions. In the first, the characteristic length of bioheat transfer $\delta_B$ is defined as

$$\delta_B = \sqrt{\frac{k}{\rho_b c_b \omega_b}} . \tag{8}$$

In the present study, three kinds of coordinates and three kinds of BCs are considered, giving nine steady-state solutions that can be derived using Eqs. (1), (4), (5), (6) and (7). In the case of the 1-D Cartesian coordinate ($n = 0$), the steady-state solutions are expressed as

$$\text{1st kind: } T_{Steady}(x) = T_B + (T_w - T_B)\exp\left(-\frac{x}{\delta_B}\right), \tag{9}$$

$$\text{2nd kind: } T_{Steady}(x) = T_B + \frac{q_w \delta_B}{k}\exp\left(-\frac{x}{\delta_B}\right), \tag{10}$$

$$\text{3rd kind: } T_{Steady}(x) = T_B + \frac{T_f - T_B}{1 + \dfrac{k}{h\delta_B}}\exp\left(-\frac{x}{\delta_B}\right). \tag{11}$$

In the case of the 1-D axisymmetric coordinate ($n = 1$), the steady-state solutions are expressed as

$$\text{1st kind: } T_{Steady}(x) = T_B + (T_w - T_B)\frac{K_0(x/\delta_B)}{K_0(x_p/\delta_B)}, \tag{12}$$

$$\text{2nd kind: } T_{Steady}(x) = T_B + \frac{q_w \delta_B}{k}\frac{K_0(x/\delta_B)}{K_1(x_p/\delta_B)}, \tag{13}$$

$$\text{3rd kind: } T_{Steady}(x) = T_B + \frac{T_f - T_B}{1 + \dfrac{k}{h\delta_B}\dfrac{K_1(x_p/\delta_B)}{K_0(x/\delta_B)}}, \tag{14}$$

where $K_n(z)$ denotes the modified Bessel function of the second order n. In the case of the 1-D spherical symmetric coordinate ($n = 2$), the steady-state solutions are expressed as

$$\text{1st kind: } T_{Steady}(x) = T_B + (T_w - T_B)\frac{x^{-1}\exp(-x/\delta_B)}{x_p^{-1}\exp(-x_p/\delta_B)}, \tag{15}$$

$$\text{2nd kind:}\quad T_{Steady}(x) = T_B + \frac{q_w \delta_B}{k} \frac{1}{1 + \frac{\delta_B}{x_p}} \frac{x^{-1}\exp(-x/\delta_B)}{x_p^{-1}\exp(-x_p/\delta_B)}, \tag{16}$$

$$\text{3rd kind:}\quad T_{Steady}(x) = T_B + \frac{T_f - T_B}{1 + \frac{k}{h\delta_B}\left(1 + \frac{\delta_B}{x_p}\right)} \frac{x^{-1}\exp(-x/\delta_B)}{x_p^{-1}\exp(-x_p/\delta_B)}. \tag{17}$$

*3.2 Definition of dimensionless parameters*

Every biological tissue or organ can be analyzed universally by introducing dimensionless parameters. By using the characteristic length of bioheat transfer $\delta_B$, dimensionless position, Fourier number (dimensionless time) and Biot number can be defined, respectively, as

$$X = \frac{x}{\delta_B}, \tag{18}$$

$$Fo = \frac{\alpha t}{\delta_B^2} = \frac{k}{\rho c}\frac{t}{\delta_B^2}, \tag{19}$$

$$Bi = \frac{h\delta_B}{k}. \tag{20}$$

To define dimensionless temperature, the steady-state temperature of biological tissue $T_B$ and steady-state surface temperature of biological tissue $T_{SS}$ are chosen as standard temperatures. The steady-state surface temperature of biological tissue $T_{SS}$ can be derived by substituting $x$ for 0 or $x_p$ in Eqs. (9)–(17). Dimensionless temperature is expressed as

$$\theta = \frac{T - T_B}{T_{SS} - T_B}. \tag{21}$$

This definition allows for the normalization of the temperature of biological tissue independent of the kind of BC.

By using these dimensionless parameters, the Pennes' bioheat transfer equation is transformed to dimensionless form as

$$\frac{\partial \theta}{\partial Fo} = \frac{1}{X^n}\frac{\partial}{\partial X}\left(X^n \frac{\partial \theta}{\partial X}\right) - \theta. \tag{22}$$

*3.3 Dimensionless steady-state solutions and BCs*

By using the dimensionless parameters given in Eqs. (18), (19), (20) and (21), nine dimensionless solutions of 1-D Cartesian, axisymmetric and spherical symmetric coordinates can be transformed to three solutions independent of the BCs:

$$\text{1-D Catesian:}\quad \theta_{Steady}(X) = \exp(-X) \tag{23}$$

$$\text{1-D Axisymmtric: } \theta_{Steady}(X) = \frac{K_0(X)}{K_0(X_p)} \tag{24}$$

$$\text{1-D Spherical symmetric: } \theta_{Steady}(X) = \frac{X^{-1}\exp(-X)}{X_p^{-1}\exp(-X_p)} \tag{25}$$

Equations (23)–(25) show that steady-state temperature distributions are independent of BCs. Figure 3 shows the temperature distributions of each coordinate; the abscissa shows the dimensionless distance from the heating/cooing surface. As shown in Fig. 3, the depth at which temperature changes in the 1-D Cartesian coordinate is the deepest of the three kinds of coordinates. Furthermore, the depth at which temperature changes in the 1-D axisymmetric and sphere symmetric coordinates decreases as the probe radius decreases.

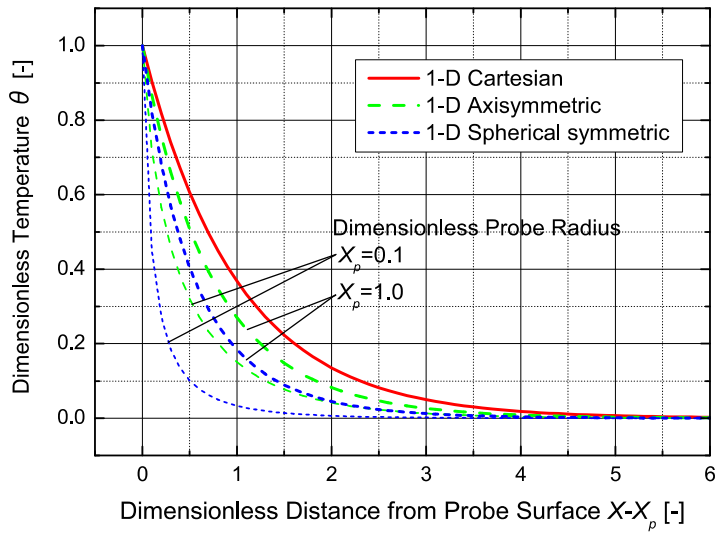

Fig. 3.  Dimensionless steady-state temperature distributions of each coordinate

Dimensionless BCs are important for transient analysis. It can be shown that BCs are independent of the coordinates by using a geometric factor $G(n)$, which is expressed as

$$G(n) = \begin{cases} 1 & (n=0;\text{Cartesian}) \\ \dfrac{K_1(X_p)}{K_0(X_p)} & (n=1;\text{Axisymmetric}) \\ 1+\dfrac{1}{X_p} & (n=2;\text{Sphere symmetric}) \end{cases}. \tag{26}$$

By using this geometric factor, dimensionless BCs are expressed as

$$\text{1st kind: } \left.\theta\right|_{Surface} = 1, \tag{27}$$

$$\text{2nd kind: } -\left.\frac{\partial \theta}{\partial X}\right|_{Surface} = G(n), \tag{28}$$

$$\text{3rd kind:} \quad -\frac{\partial \theta}{\partial X}\bigg|_{Surface} = Bi\left[\left(1 + \frac{1}{Bi}G(n)\right) - \theta\big|_{Surface}\right]. \tag{29}$$

Furthermore, the steady-state surface temperatures of biological tissue $T_{SS}$ can be expressed independently of coordinates by the geometric factor

$$T_{SS} = \begin{cases} T_w & \text{(1st kind)} \\ T_B + \dfrac{q_w \delta_B}{k}\dfrac{1}{G(n)} & \text{(2nd kind)} \\ T_B + \dfrac{T_f - T_B}{1 + \dfrac{k}{h\delta_B}G(n)} & \text{(3rd kind)} \end{cases}. \tag{30}$$

Figure 4 shows the variation of geometric factors of each radius for the dimensionless probe radius. As shown in Fig. 4, all geometric factors converge to 1 as the dimensionless probe radius gets larger. This means that every coordinate can be approximated to 1-D Cartesian coordinates when the dimensionless probe radius is large enough. By using the data shown in Fig. 4, the BCs and the steady-state surface temperature can be calculated for the arbitral probe radius from Eqs. (27)–(30).

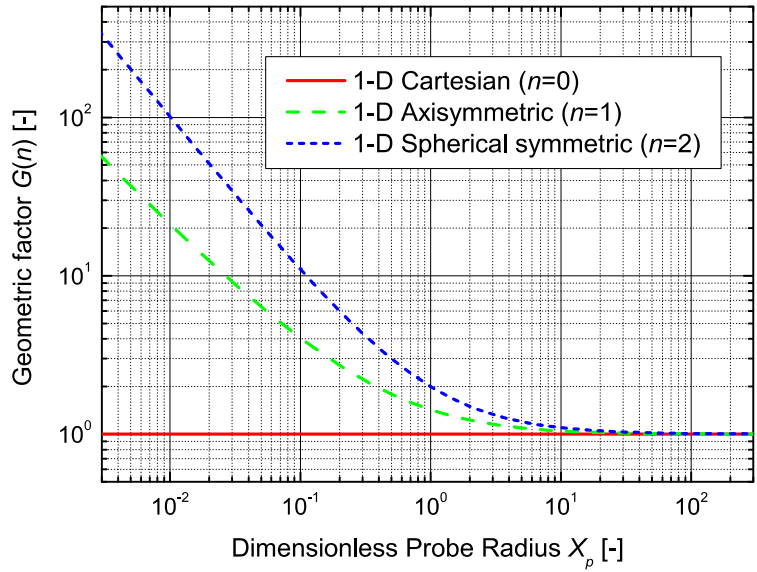

Fig. 4.  Geometric factors of each coordinate

## 4. General Characteristics of Bioheat Transfer

*4.1 Steady-state thermal penetration depth*

A steady-state in temperature distribution in a biological tissue indicates the existence of a limit depth where temperature varies. This depth is an important parameter in estimating the depth affected by heat transfer during treatment. In the present study, the steady-state thermal penetration depth (SSTPD) $\delta_T$ is defined as the distance from the heating/cooling surface to the point at which the dimensionless temperature is 10%. This threshold is adopted because uncertainties in biological tissues, such as the distribution of blood vessels and differences between individuals, should be taken into consideration in any practical estimation of temperature distribution. Dimensionless SSTPDs $\Delta_T$ in each coordinate are expressed as

1-D Cartesian: $\quad \Delta_T = \ln 10 = 2.30$, (31)

1-D Axisymmetric: $\quad K_0(\Delta_T + X_p) = 0.1 K_0(X_p)$, (32)

1-D Spherical symmetric: $\quad \dfrac{\exp(\Delta_T + X_p)}{\Delta_T + X_p} = 0.1 \dfrac{\exp(X_p)}{X_p}$. (33)

It is important to note that the SSTPD is independent of the BC and depends only on the dimensionless probe radius $X_p$. The 1-D Cartesian SSTPD thus attains a constant value. Figure 5 shows the SSTPDs for the dimensionless probe radius. As shown in Fig. 5, the 1-D spherical symmetric SSPTD is the smallest in all three kinds of coordinate. Furthermore, the 1-D axisymmetric and spherical symmetric SSTPDs converge at 2.30, which is the value of the SSTPD of the 1-D Cartesian coordinate as the dimensionless probe radius becomes larger. These results are described in a dimensionless form and can therefore be applied to any biological tissue or organ by using the follow definition:

$$\delta_T = \Delta_T \times \delta_B = \Delta_T \sqrt{\dfrac{k}{\rho_b c_b \omega_b}} \ . \tag{34}$$

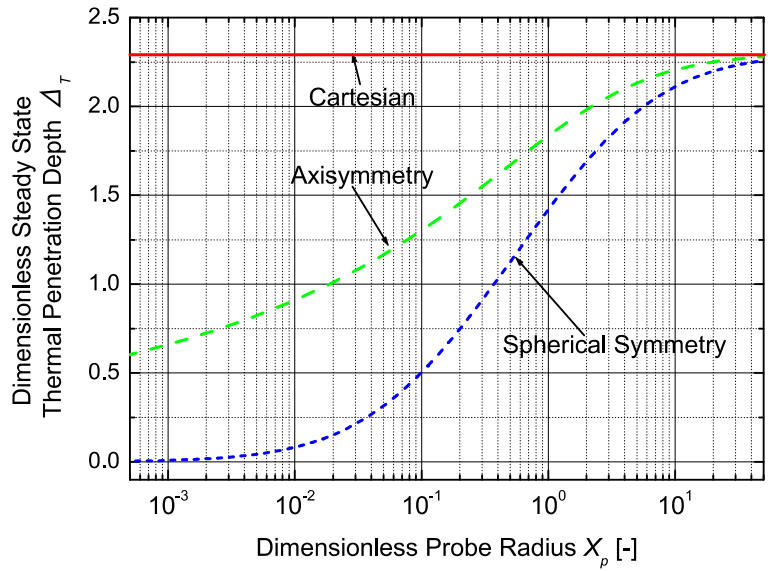

Fig. 5. Dimensionless steady-state thermal penetration depths of each coordinate

*4.2 Time to reach steady-state*

The time to reach steady-state temperature distribution in a biological tissue is an important parameter in the estimation of treatment time. First, the relative difference between steady-state and transient solution is defined as

$$RD[\%] = \dfrac{1}{\Delta_T} \int_{X_p}^{\Delta_T + X_p} \dfrac{\theta_{Steady}(X) - \theta_{Transient}(Fo, X)}{\theta_{Steady}(X)} dX \times 100 \ . \tag{35}$$

In the case of the 1-D Cartesian coordinate, Eq. (35) is valid by substituting 0 for $X_p$. The integral range is from the tissue surface to the SSTPD. The relative difference indicates the level of time evolution of the temperature distribution. The initial and steady states are represented by 100% and 0%, respectively. It is difficult to determine a precise steady-state. Therefore the time to reach steady-state is defined as the time when the relative difference becomes 10%. Dimensionless transient solutions are solved only by numerical simulation in the present study because analytical dimensionless transient solutions $\theta_{Transient}$ are difficult to derive. Equation (22) is discretized by the finite volume method and the perfect implicit method.

The time to reach steady-state depends on the BC and coordinates. Figure 6 shows the dimensionless time in the 1-D Cartesian coordinate. In the case of the 1-D Cartesian coordinate and third kind BC, the dimensionless time to reach steady-state depends only on the Biot number. In addition, the time to reach steady-state of first kind BC 1.35 is the minimum value, while that of second kind BC 2.30 is the maximum value. The time of third kind BC varies from 1.35 to 2.30. Figures 7 and 8 show the time to reach steady-state of the 1-D axisymmetric and spherical symmetric coordinates, respectively. As shown in the two figures, the minimum and maximum times are the values in the case of first and second kind BCs, respectively. The time of 1-D spherical symmetry is smaller than that of axisymmetry. This means that the temperature distribution of 1-D spherical symmetry has a smaller size and quicker convergence than the other coordinates. These results are described in a dimensionless form and can therefore be applied to any biological tissues or organs by using the follow definition:

$$t_{RSS} = Fo_{RSS} \frac{\delta_B^2}{\alpha} = Fo_{RSS} \delta_B^2 \frac{\rho c}{k}. \tag{36}$$

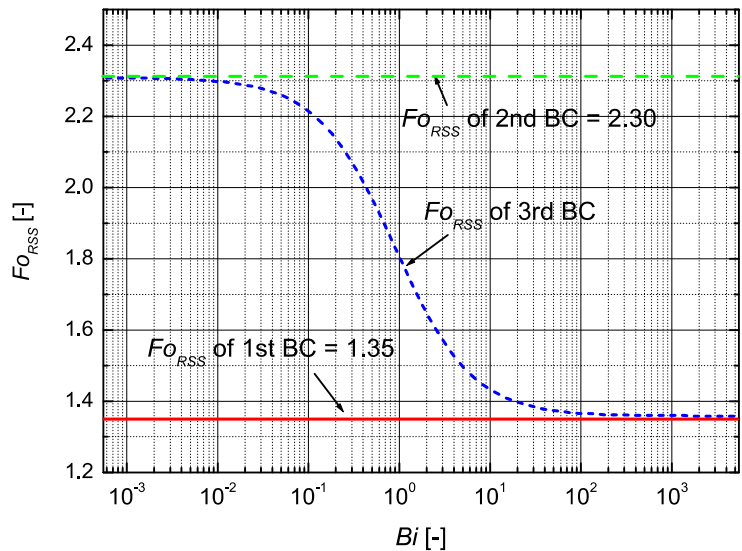

Fig. 6. Dimensionless time to reach steady-state in 1-D Cartesian coordinate

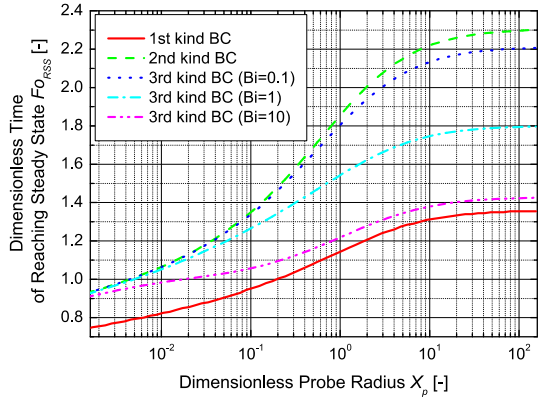
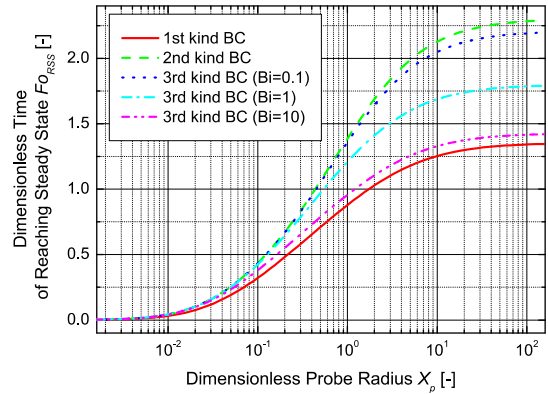

Fig. 7. Dimensionless time to reach steady-state of 1-D axisymmetric coordinate

Fig. 8. Dimensionless time to reach steady-state of 1-D spherical symmetric

## 5. Application

The results derived in the present study can be applied to thermal problems in several biological tissues universally because of the dimensionless form. During thermal therapy, it is difficult to estimate the characteristics of temperature distribution in the biological tissue quantitatively and completely because of individual differences in the human body. However, the results in the present study are independent of individual differences and can therefore be used in estimations for thermal therapy. The typical thermophysical properties of biological tissue and organs and the calculated characteristic length of bioheat transfer are listed in Table 1. Table 2 shows the SSTPD and the time to reach steady-state of the 1-D Cartesian coordinate. Internal organs have a high blood perfusion rate, and the SSTPD of the internal organs is therefore small; the SSTPD of body surface tissue (skin, adipose tissue and skeletal muscle) is larger than that of the internal organs. Table 2 also shows that the maximum depth reached by the heat effect is approximately 3cm. Meanwhile, the time needed to reach a steady-state varies in accordance with the blood perfusion rate except in the adipose tissues, where it takes longer than in other biological tissues because the thermal diffusivity of adipose tissue is smaller. Table 2 provides guidelines about the extent of temperature distribution and the time for treatment. The solutions and equations derived in the present study are summarized in Table 3. Temperature in biological tissue can be estimated by providing the geometry of the treatment (coordinate and probe radius) using Table 3.

Table 1. Thermophysical properties of each organ or tissue

| | Thermal conductivity $k$ [W/(K·m)] | Specific heat $c$ [J/(kg·K)] | Density $\rho$ [kg/m$^3$] | Blood perfusion rate $\omega_b$ [1/s] | Characteristic length of bioheat transfer $\delta_B$ [mm] |
|---|---|---|---|---|---|
| Kidney[*1] | 0.54 | 3700 | 1050 | 6.1×10$^{-2}$ | 1.5 |
| Liver[*1] | 0.52 | 3600 | 1060 | 1.5×10$^{-2}$ | 3.0 |
| Prostate[*2] | 0.50 | 3850 | 1000 | 6.1×10$^{-3}$ | 4.5 |
| Brain(White matter)[*3] | 0.50 | 3700 | 1050 | 3.5×10$^{-3}$ | 6.0 |
| Skin(Dermis)[*4] | 0.45 | 3300 | 1200 | 1.3×10$^{-3}$ | 9.6 |
| Adipose tissue[*1] | 0.27 | 3100 | 950 | 5.0×10$^{-4}$ | 11.7 |
| Skeletal muscle[*1] | 0.50 | 3465 | 1050 | 9.0×10$^{-4}$ | 11.9 |
| Blood[*1] | 0.51 | 3720 | 1060 | - | - |

*1：Brix et al., 2002. *2：Zhang et al., 2005. *3：Diao et al., 2003. *4：Jiang et al., 2002.

Table 2. Steady-state thermal penetration depth and time to reach steady-state of 1-D Cartesian coordinate

|  | Steady-state thermal penetration depth $\delta_T$ [mm] | Time to reach steady-state (BC of the 1st kind) $t_{RSS}$ [s] | Time to reach steady-state (BC of the 2nd kind) $t_{RSS}$ [s] |
|---|---|---|---|
| Kidney | 3.4 | 22 | 37 |
| Liver | 6.8 | 87 | 148 |
| Prostate | 10.5 | 215 | 366 |
| Brain(White matter) | 13.8 | 380 | 647 |
| Skin(Dermis) | 22.0 | 1084 | 1847 |
| Adipose tissue | 26.9 | 2016 | 3435 |
| Skeletal muscle | 27.3 | 1384 | 2357 |

Table 3. The summary of the solutions of bioheat transfer equation and BCs

| Coordinate system | Dimensionless Steady-state solution $\theta_{steady}(X)$ | Geometric Factor $G(n)$ | B.C. of 1st kind | | B.C. of 2nd kind | | B.C. of 3rd kind | |
|---|---|---|---|---|---|---|---|---|
| | | | Dimensionless form | Steady-state surface temperature | Dimensionless form | Steady-state surface temperature | Dimensionless form | Steady-state surface temperature |
| Cartesian ($n$=0) | $\exp(-X)$ | 1 | $\theta\|_{Surface} = 1$ | $T_w$ | $-\dfrac{\partial \theta}{\partial X}\bigg\|_{Surface} = G(n)$ | $T_B + \dfrac{q_w \delta_B}{k}\dfrac{1}{G(n)}$ | $-\dfrac{\partial \theta}{\partial X}\bigg\|_{Surface} = Bi\left[\left(1+\dfrac{1}{Bi}G(n)\right) - \theta\|_{Surface}\right]$ | $T_B + \dfrac{T_f - T_B}{1+\dfrac{k}{h\delta_B}G(n)}$ |
| Axisymmetric ($n$=1) | $\dfrac{K_0(X)}{K_0(X_p)}$ | $\dfrac{K_1(X_p)}{K_0(X_p)}$ | | | | | | |
| Sphere symmetric ($n$=2) | $\dfrac{X^{-1}\exp(-X)}{X_p^{-1}\exp(-X_p)}$ | $1+\dfrac{1}{X_p}$ | | | | | | |

Two examples of calculations of bioheat transfer characteristics by using Tables 1–3 are shown. The problem of hyperthermia using a needle-type probe in the liver is solved as the first example. If the heating section of the probe is limited to the tip, the 1-D spherical coordinate is a good approximation. Furthermore, the approximation of the second kind BC is valid if the probe is heated by Joule heating. When the probe radius is 3mm, the dimensionless probe radius $X_p$ is 1.0 from Table 1 and Eq. (2). In this case, the dimensionless SSTPD and dimensionless time to reach steady-state are 1.4 and 1.3 from Figs. 5 and 6, respectively. Therefore the SSTPD and time to reach steady-state are 4.2 mm and 85 s based on Eqs. (34) and (36). These values are independent of the magnitude of heat flux, and the geometric factor is 2.0; therefore, the surface temperature of the probe can be estimated as a function of heat flux by using Eq. (30). The problem of skin surface heating using the infrared radiation device we developed (Ogasawara et al., 2008) is solved as the second example. In this case, the 1-D Cartesian coordinates and second kind BC are suitable. The characteristic length of bioheat transfer of the skin is 9.6 mm according to Table 1. Furthermore, the SSTPD and dimensionless time to reach steady-state are 22 mm and 1847 s, respectively, from Table 2. Using these values, the surface temperature of the skin can be estimated from Eqs. (26) and (30).

# 6. Conclusion

Novel notations of the dimensionless solutions of the bioheat transfer equation and bioheat transfer characteristics are proposed in the present study.

- Our new definition of dimensionless temperature based on the bioheat transfer characteristics eliminates differences in BCs in steady-state solutions.
- Steady-state thermal penetration depths can be obtained. These parameters are independent of the kind of BC; therefore, the depth at which temperature changes can be described regardless of the heating temperature or heating method.
- The dimensionless time to reach steady-state is calculated. The time becomes the maximum and minimum in the case of the second and first kind BC, respectively.

One-dimensional coordinate systems were only discussed. However results in this paper can be applicable to the 2- or 3-D coordinate systems when the size of the heating region is large enough. The important point of the present study is that all data are given in a dimensionless form. Therefore these results provide the comprehensive and general characteristics of the bioheat transfer during the various thermal therapies. This makes it possible for medical doctors in the clinical field to predict temperature distribution in the human body by using the present results together with the in-situ measurement of the blood perfusion rate.

## Nomenclature

| | | |
|---|---|---|
| $Bi$ : | | Biot number, - |
| $c$ : | | specific heat, J/(kg·K) |
| $Fo$ : | | Fourier number, - |
| $Fo_{RSS}$ : | | dimensionless time to reach steady-state, - |
| $G$ : | | Geometric factor, - |
| $h$ : | | heat transfer coefficient, W/(K·m$^2$) |
| $k$ : | | thermal conductivity, W/(K·m) |
| $n$ : | | the number to classify coordinates, - |
| $q$ : | | heat flux, W/m$^2$ |
| $q_{met}$: | | metabolic heat generation rate, W/m$^3$ |
| $RD$ : | | relative difference, % |
| $T$ : | | temperature, °C |
| $T_B$ : | | steady-state temperature of biological tissue, °C |
| $T_{SS}$ : | | steady-state surface temperature of biological tissue, °C |
| $t$ : | | time, s |
| $t_{RRS}$ : | | time to reach steady-state, s |
| $X$ : | | dimensionless position, - |
| $x$ : | | position, m |

Greek

| | | |
|---|---|---|
| $\alpha$ : | | thermal diffusivity, m$^2$/s |
| $\Delta_T$ : | | dimensionless steady-state thermal penetration depth, m |
| $\delta_B$ : | | characteristic length of bioheat transfer, m |
| $\delta_T$ : | | steady-state thermal penetration depth, m |
| $\theta$ : | | dimensionless temperature, - |
| $\rho$ : | | density, kg/m$^3$ |
| $\omega_b$ : | | blood perfusion rate, 1/s |

Subscripts

| | | |
|---|---|---|
| *a* : | | artery |
| *b* : | | blood |
| *f* : | | ambient |
| *p* : | | probe |
| *steady* : | | steady-state |
| *transient* : | | transient |
| *w* : | | surface of biological tissue |

**Acknowledgements**

This work was supported by a Grant-in-Aid for Scientific Research (A) [18206022] from the Ministry of Education, Culture, Sports, Science, and Technology, Japan, and by a Grant-in-Aid for JSPS Fellows [20·7374] from the Japan Society for the Promotion of Science.